\documentclass[preprint2]{aastex}
\shorttitle{Water in V838 Mon}
\shortauthors{Banerjee, Barber, Ashok $\&$ Tennyson}
\begin{document}
\title{Near-Infrared water lines in V838 Monocerotis}

\author{D.P.K. Banerjee}
\affil{Physical Research Laboratory, Navrangpura,  Ahmedabad 
Gujarat 380009, India}
\email{orion@prl.ernet.in}
\author{R.J. Barber}
\affil{Department of Physics and Astronomy, University College London, London WC1E 6BT, UK}
\email{bob@theory.phys.ucl.ac.uk}

\author{N.M. Ashok}
\affil{Physical Research Laboratory, Navrangpura,  Ahmedabad 
Gujarat 380009, India}
\email{ashok@prl.ernet.in}
\and
\author{J. Tennyson}
\affil{Department of Physics and Astronomy, University College London, London WC1E 6BT, UK 
}
\email{j.tennyson@ucl.ac.uk}

\begin{abstract}
V838 Monocerotis had an intriguing, nova-like outburst in January 2002 which has
subsequently led to several studies of the object. It is now recognized that
the  outburst of V838 Mon and its  evolution  are different
from that of a classical nova or other classes of well-known eruptive variables. 
V838 Mon, along with two other objects that have analogous
properties, appears to comprise a new class of eruptive variables. There
are limited infrared studies of V838 Mon. Here, we present near-infrared $H$ band
 (1.5 - 1.75$\mu$m) spectra of V838 Mon from late 2002 to the end of 2004. The
principal, new result from our work is the detection of several,
 rotation-vibration lines of water in the $H$ band spectra. The observed water lines 
have been modeled  to first establish that they are indeed due to water. Subsequently  the 
temperature and column densities of the absorbing material, from
where the water absorption features originate,  are derived. From our analysis,
we find that the water features arise from a cool $\sim$ 750-900 K region around 
V838 Mon which appears to be gradually cooling with time.  
\end{abstract}

\keywords{infrared: stars-novae, cataclysmic variables - stars: individual 
(V838 Monocerotis) - techniques: spectroscopic}

\section{Introduction}
Water vapor is the most important gas phase molecule in astrophysical
environments after H$_{2}$ and CO. It is an abundant molecule
in the atmospheres of cool, oxygen rich (O/C $>$ 1) stars and
it constitutes the major opacity source in the infrared in such stars. The
detection of water vapor, through clearly identified spectral lines, whilst not
a common event  has been made in a variety of sources. Some of the previous 
detections
are in the photospheres of K and 
M giants/supergiants (e.g. $\alpha$ Ori and $\alpha$ Sco - Jennings $\&$ Sada 1998;
 $\alpha$ Boo - Ryde et al. 2002;  $\alpha$ Tau and  other stars - Tsuji, 2001); in
  sunspots (Wallace et al. 1995, Polyansky et al. 1997); in young stellar objects (Carr, Tokunaga $\&$ 
  Najita 2004) and in comets (Crovisier et al.1997, Dello Russo et al. 2005 ).
Detection in other sources such as Herbig-Haro objects, star forming regions,
planetary atmospheres and in outflows from evolved stars like VY CMa are
described in Neufeld et al (1999; and references therein). From a survey of
the literature, it  appears that the detection of water lines in novae or 
nova-like variables is either rare or has not been made earlier.
Most of the water detections described above are in the infrared region at 
wavelengths beyond the $K$ band (i.e. beyond 2.4 $\mu$m).
Several of them have  been made by the Infrared Space Observatory (ISO) from
spectra obtained in the 2.48-40 $\mu$m region. In  the $JHK$ (1.0 to 2.4 $\mu$m) 
region or its vicinity, there are limited reports of water line detections. In 
this context we note the  detection by Spinrad 
and Newburn (1965) of a water band at $\sim$  9400A in the spectra of cool M
type stars and also the detection of water bands in the Mira variable R Leo by
Hinkle and Barnes (1979) from high-resolution spectra.

V838 Mon was reported in eruption on 2002 January 6 by Brown (2002).
 This initial outburst had  a peak  magnitude of V$_{max}$ = 10. Subsequent 
observations showed two more outbursts  in the object (reaching a V$_{max}$ of 6.7 
and 7 respectively) within the next two months. V838 Mon also showed a fast cooling
 to a cool, late M spectral type. The multi-peaked
lightcurve and the decrease of the effective temperature with time suggested that 
the object was different from a classical nova. The early photometric and 
spectroscopic observations are described in Munari et al. (2002), Kimeswenger et al.
2002), Crause et al. (2003) and Wisniewski et al. (2003). An expanding light-echo was
also seen around the star (Henden,  Munari $\&$ Schwartz 2002, Bond et al. 2003). 
Estimates  based on the expanding light echo, suggest that the object is at a
large distance, in the 8-10 kpc range (Tylenda 2004, Crause et al. 2005, van Loon et al. 2004). 
There is some debate about whether the scattering medium causing the light echo is an 
interstellar sheet of dust in the line of sight or matter lost in an earlier episode 
from the progenitor of V838 Mon - which would thereby imply it to be an evolved red 
giant or AGB star  (van Loon et al. 2004; Tylenda, Soker $\&$ Szczerba 2005). Several 
suggestions have been made, that V838 Mon  has similar  properties as V4332 Sgr and 
M31-RV (two other objects which have erupted 
in the last 10-15 years) and that these objects taken together could form a new 
class of eruptive variables. Our recent results on V4332 Sgr certainly show it to be 
an unusual object (Banerjee $\&$ Ashok 2004, Banerjee et al. 2004). 
Since the outburst properties of V838 Mon are different from known eruptive variables,
 new mechanisms have been proposed for the outburst 
in such objects. Soker $\&$ Tylenda (2003) propose a merger between main 
sequence stars while Retter $\&$ Marom (2003) suggest a planetary capture for the 
eruption. In the case of V4332Sgr specifically, observations suggest (Banerjee et 
al., 2004) that the second mechanism could be plausible.  V838 Mon has been the 
subject of several other studies viz. polarization studies (Desidera et al. 2004, 
Wisniewski, Bjorkman $\&$  Magalhaes 2003), elemental abundance determinations 
(Kipper et al. 2004, Kaminsky $\&$ Pavlenko 2005), interferometric measurements to 
determine the size of the object (Lane et al. 2005) and investigations of the 
progenitor of V838 Mon (Munari et al. 2005, Tylenda et al. 2005). However many 
aspects of V838 Mon still remain unclear, including the definitive cause for its 
eruption.

There are two studies of V838 Mon in the infrared which are specially relevant 
to the present  work. Evans at al. (2003), based on an IR spectrum obtained in 
October 2002 in the 0.8-2.5$\mu$m region, showed that the spectrum of V838 Mon 
has similarities to a  cool L type dwarf. Lynch et al (2004), using 
multi-epoch IR spectra between January 2002 to March 2003,  computed a 
detailed model of V838 Mon and its circumstellar environment. 
Their model indicates a cool central 
star with a photospheric temperature of $\sim$ 2100 K surrounded by a large,
spherical envelope of molecular matter at $\sim$ 750-800 K. In both these studies, 
the authors point out the presence of water in V838 Mon (see Figs. 1 $\&$ 2 
in Evans et al. 2003 and
Figs. 6 $\&$ 7 in Lynch et al. 2004). In these figures, the presence of water 
is indicated  
through the wide and deep absorption troughs  seen between the spectral bands 
e.g at 
1.4 ${\rm{\mu}}$m between $J$ $\&$ $H$ bands and at 1.9${\rm{\mu}}$m between
the $H$ 
$\&$ $K$ bands. Such inter-band, water features are well known and commonly 
observed 
in cool M stars (e.g. see the IR catalog of stellar spectra by  
Lancon and Rocca-Volmerange
1992) and also in brown dwarfs (e.g. the spectral catalog by Geballe et 
al.2002). In addition, Figure 12 of Lynch et al. (2004) shows water lines
in the $K$ band in both their model and observed data. However, the studies by Evans et al. (2004) and Lynch et al. (2004) 
do not identify  and label specific water lines as we do here. 
Furthermore, our spectra  extend to late 2004, beyond the 
observations reported in the above studies.

\section{Observations}
 Since its outburst in early 2002, near-IR $JHK$ spectra of V838 Mon have been
 taken, until the present day,  at approximately  bi-monthly intervals from the 
 1.2m telescope at the Mt. Abu Observatory.  Observations from mid-June to early 
 October are not possible due to the rainy season. Spectra early after the outburst 
 until May 2002) are presented in Banerjee $\&$ Ashok (2002).
 The $H$ band spectra presented here cover the period from end-2002 to 
 end-2004 and are taken at fairly equi-spaced epochs - they are thus suitable to study the 
 evolution of the  object. The present spectra, like those of our earlier study,
 were obtained at a resolution of $\sim$ 1000 using a Near Infrared Imager/Spectrometer
 with a 256$\times$256 HgCdTe NICMOS3 array.  Generally, a set of at least two spectra
  were taken  with the  object dithered to  two 
 different positions along the  slit.   The spectrum of the comparison star  HR 2714 
 (A2V spectral type), after removing the hydrogen lines in its spectrum,  was used
 to ratio the spectra of V838 Mon. HR 2714 is spatially close to V838 Mon ($\sim$ 
 3 degrees away) and furthermore its  spectra were  recorded immediately preceding/after 
 the V838 Mon observations. Thereby it is ensured that the object and the comparison 
 star are  observed at similar air-mass. The ratioing process should therefore reliably 
 remove  telluric lines  in the V838 Mon spectra. Wavelength calibration   was done 
 using OH sky lines  that register with the spectra. Spectral reduction and analysis
 were done  using IRAF. The observational details are presented in Table 1.

\section{Results}
The H band spectra are shown in Figure 1. Apart, from several prominent absorption 
features beyond 1.73$\mu$m which we establish below to be due to water, there are other
strong features in the spectra  attributable to AlO and the second overtone of $^{12}$CO. 
We briefly discuss these features here but a more comprehensive modeling and  analysis 
of their evolution, supplemented by our simultaneous   $J$ and $K$ band spectra in which
these molecules have strong spectral signatures, will be presented in a future work. This 
work will also incorporate our $JHK$ photometry results between 2002-2004. The most 
prominent features in Fig. 1 are the deep absorption features at 1.6480${\rm{\mu}}$m  and
1.6837 ${\rm{\mu}}$m  which are due to the (1,0) vibronic transitions from the A-X  band 
system of the  AlO radical. Along with its possible analog V4332 Sgr,  V838 Mon appears to
be the only other object to show these rare AlO bands (Banerjee et al. 2003). AlO  has a 
strong (4,0) A-X band in the $J$ band also which can be seen
in the spectra of Lynch et al. (2004) and Evans et al. (2003). The expected 
positions of the $^{12}$CO first overtone bands ($\Delta$$\nu$ = 3) are marked in Figure 
1. Transitions from 3-0 to 7-4 are fairly prominent, especially in
the later spectra. In general, the CO emission in V838 Mon is rather complex, especially 
in the first overtone bands starting from 2.29${\rm{\mu}}$m in the $K$
band. These first overtone CO overtone bands are unusually deep in V838 Mon and show a 
very complex structure and evolution (Banerjee $\&$ Ashok 2002, Lynch et al. 2004, Evans 
et al. 2003 ). An integrated modeling, using both $H$ and $K$ band spectra, is necessary
to understand the behavior of CO in V838 Mon.

 \subsection{Synthetic spectra of water}
Our focus, in this study, is on the presence of water in V838 Mon. We tested for the
signature of water in the spectra using the BT2 line list. BT2 (Barber et al 2005) is a variational line list, computed at University College London using the DVR3D suite
of programs that calculates the rotation-vibration spectra of triatomic
molecules (Tennyson et al., 2004). It is the most complete water line list in
existence, comprising over 500 million transitions (65\% more than
any other list) and is also the most accurate (over 90\% of all known
experimental energy levels are within 0.3 cm$^{-1}$ of the BT2 values). 
The final module of the DVR3D suite, `SPECTRA', was used to generate
synthetic spectra at the appropriate wavelengths and temperatures,
the output being convolved at 4 cm$^{-1}$ (similar to the resolving
power of the instrument). Finally the output was `binned' with the
same bin sizes and positions as the observed spectrum.
Water features were clearly identified. Moreover, the relative intensities
of the various features in the synthetic spectra are  temperature-dependent,
and by testing for a best fit with the observed spectra, it was possible
to determine the temperature of the region within which absorption
was taking place on each of the five dates (Figure 2). Fig. 2 shows the
best fit temperature on each date. The observed spectra are shown
as histograms, whilst the synthetic spectra are presented in their
un-binned form. It will be seen from Fig. 2 that between 20 November 2002 
and 25 December 2004 the temperature at which absorption occurred decreased 
from 900$\pm$30 K to 750$\pm$50 K, that is to say a reduction of 150$\pm$60 K over a period of 767 days.

Many of the transitions within the BT2 line list have been labeled,
and consequently it was possible to assign the 17 strong lines that
comprise the five main absorption features in Fig. 2. The details
are given in Table 2. The lower and upper levels for each of these
transitions are labeled in the manner: ($\nu_{1}$ $\nu_{2}$ $\nu_{3}$){[}J
K$_{\textrm{a}}$K$_{\textrm{c}}${]}, where the terms in round brackets
are the vibrational quantum numbers, those in the square brackets represent
the asymmetric top rotational quantum number and its projection onto
two orthogonal axes (A and C) respectively. It will be seen that with
the exception of one line at 1.74454 $\mu$m all of these strong transitions
are in the (0 0 0)-(0 1 1) band. Table 2 also contains information about 
intensity. The fourth column
gives the relative intensities of the 17 strongest synthetic H$_{2}$O
lines in the wavelength range 1.726 to 1.751 $\mu$m at a temperature
of 800 K, whilst the last column gives the relative intensity of the
five strong absorption features that are the result of the blending
of these lines. The intensity is expressed as the total integrated
line intensity within the bin whose position corresponds to that of
maximum absorption within the feature. The apparent discrepancies
between the two sets of data are explained by the fact that the features
comprise not only the strong lines listed in Table 2, but a large
number of weak lines which, individually, are insignificant, but collectively
can make an important contribution to total absorption.

Lastly, it has been possible to estimate the H$_{2}$O column densities
using $I=I_{0}$exp$(-\kappa_{\lambda}NS)$, where N is the number of water
molecules per cm, S is the distance in cm and the product NS is the
column density (molecules cm$^{-1}$). $\kappa_{\lambda}$ is the
opacity at wavelength $\lambda$ and is one of the outputs of SPECTRA.
Once the best temperature fit had been established, the optimum values
of I$_{0}$ and NS which gave the best fit to the observed data were
obtained. The effect of increasing I$_{0}$ is to raise the overall
level of the synthetic plot, whilst increasing NS had the effect of
increasing the depth of the strong absorption features relative to
the weak ones. The data used for this fitting, prior to applying the
exponential factor above,  was first convolved with a gaussian 
with a FWHM of 1 cm$^{-1}$ which is equivalent to a Doppler velocity spread
of 50 km/s. Such a velocity spread in the cool expanding shell around V838 Mon 
is suggested from the analysis of Lynch 
et al. (2004). Subsequently the model data was convolved with a 4 cm$^{-1}$ FWHM gaussian to take account of the resolving power of the instrument. Table 3 details our estimates of column density on each of the five
dates. We believe that our methodology is liable to suffer from systematic
errors and consequently we estimate the errors as being +100\%, -50\%.
However, the comparison between the column densities at the different
dates will be more accurate, and we estimate the error in the relative
numbers at +25\%, -20\%.

\section{Discussion}
We have compared our column density and temperature estimates with those
derived by Lynch et al. (2004). The temperature of 800$\pm$30 K on 25 January 
2003 is consistent with the 750-790 K suggested by Lynch et al. (2004)
for the `cool cloud' surrounding V838 Mon based on observations of
H$_{2}$O and CO molecular bands and the SiO (2$\nu$) feature made
at about the same date. Our estimate of the H$_{2}$O column density
on 25 January 2003 is $\sim$ 63$\%$ of the Lynch et al. (2004) value at a
similar date and is therefore in reasonably good agreement (within our estimated
error) with their work. The presence of an envelope at $\sim$ 800 K  around 
V838 Mon, as our analysis shows, confirms the findings of Lynch et al. (2004) 
that a cool cloud surrounds the central star of V838 Mon. The derived temperature
 of 800K is close to that 
expected for an expanding grey body which absorbs stellar energy in the VIS-NIR 
and reradiates it at longer wavelengths (see equation 4, Lynch et al. 2004. If 
the cooling of the water-bearing envelope continues at the suggested rate 
of $\sim$ 75 K per 
year, its temperature may reach the ice sublimation temperature ($\sim$ 150 K) 
to form water-ice in a few years from now. Such a development would further 
enhance the similarity between V838 Mon and its possible analog V4332 Sgr in 
which water  ice is detected strongly 10 years after the object's 
outburst (Banerjee et al. 2004). 
 
\begin{acknowledgements}
Research work at the Physical Research Laboratory  is funded by the Department of 
Space, Government of India. R.J. Barber wishes to thank the Particle Physics 
and Astronomy Research Council for its financial support. We also thank the
anonymous referee for his helpful comments.  	  
	  
\end{acknowledgements}


\clearpage
\begin{figure}
\plotone{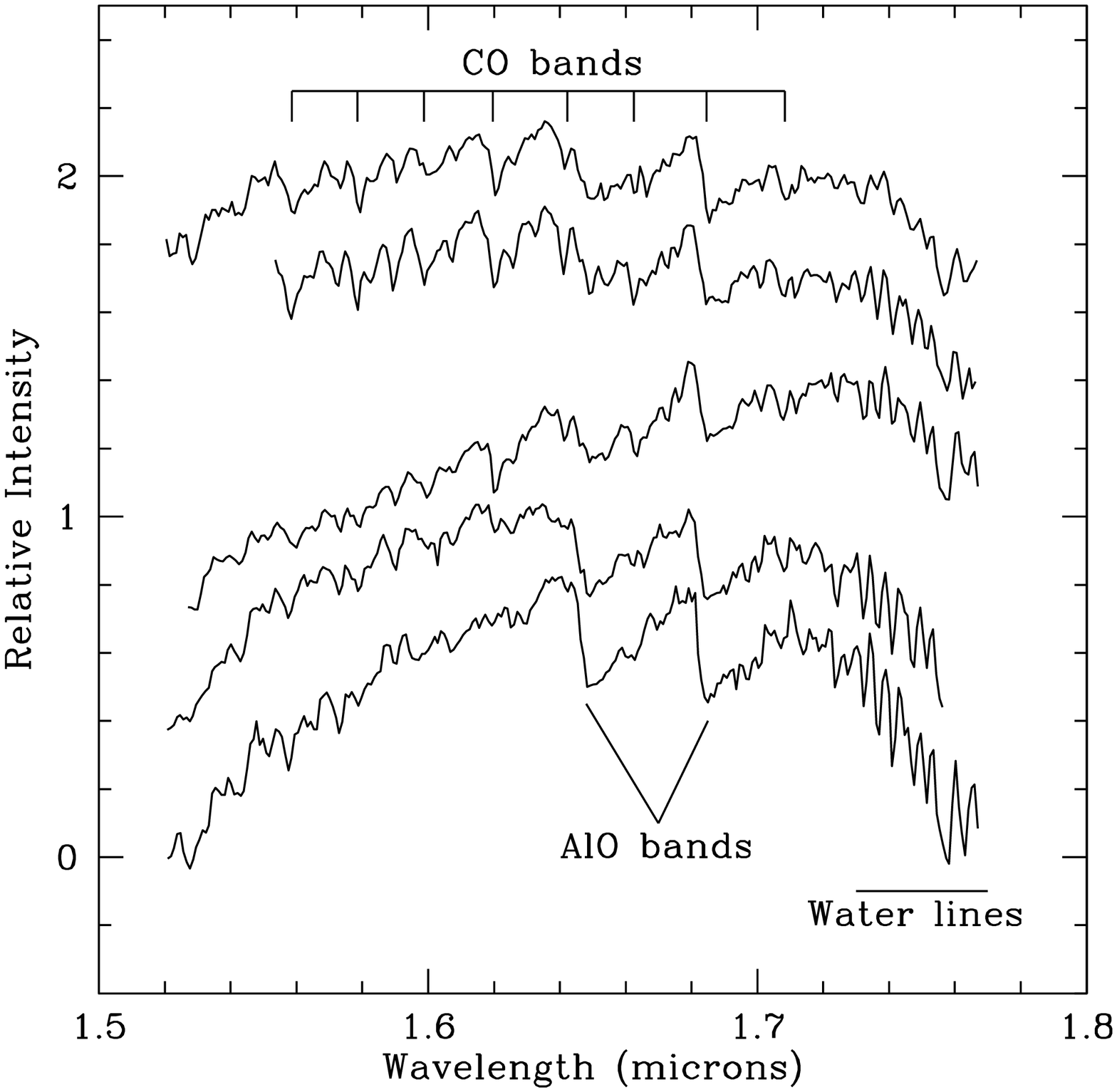}
\caption{The near-IR $H$ band spectra of V838 Mon are shown, from bottom up, 
for 20 Nov 02, 25 Jan 03, 14 Dec 03, 15 Apr 04 and 25 Dec 04. The prominent water 
lines are marked. Also shown are the positions of the second overtone ($\Delta$$\nu$ = 3)
bandheads of $^{12}$CO.  The prominent (1,0) AlO molecular bands arising from the 
$A$$^{\rm 2}$$\Pi$$_{i}$$-$$X$$^{\rm 2}$$\Sigma$$^{\rm +}$ band system can be 
seen. The spectra have been offset by arbitrary amounts in intensity units for clarity
of presentation. However, the true  continuum strength at the $H$ band center 
(1.65${\rm{\mu}}$m) can be estimated from broad band photometric fluxes available 
on days close to our observations (Crause et al. 2005). For the spectra from bottom 
to top, these are  $H$ = 5.31, 5.53, 5.87, 5.92 and 6.18 mag. respectively. 
\label{fig1}}
\end{figure}

\clearpage
\begin{figure}
\vspace{-20mm}
\plotone{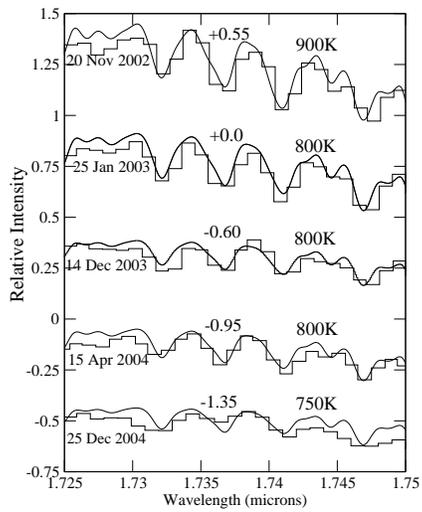}
\caption{The synthetic spectra (continuous lines) superposed on the observed
data (in binned form) for the  different epochs of observations. The observed
spectra were normalized to unity at 1.65$\mu$m and for clarity are offset with 
the additive constants indicated above each spectra.   
\label{fig2}}
\end{figure}

\clearpage
\begin{table}
 \caption{Log of  observations for V838 Mon. }
\begin{tabular}{ccccc}
\hline 
\hline\\
Date $\&$ UT         &Integration time (s) &   Airmass & Std. star &  Airmass    \\
                     &      V838 Mon       &  V838 Mon &           &Standard star\\
20 Nov 2002(0.9056)  &       20            &   1.114   & HR 2714   &   1.148     \\
25 Jan 2003(0.8000)  &       45            &   1.203   &    "      &   1.161     \\
14 Dec 2003(0.8771)  &       40            &   1.144   &    "      &   1.109     \\
15 Apr 2004(0.6542)  &       30            &   1.655   &    "      &   1.567     \\
25 Dec 2004(0.8250)  &       60            &   1.150   &    "      &   1.080     \\

\hline
\hline
\hline\\
\end{tabular} 
\end{table}

\clearpage
\begin{table}
\caption{Line Center and Relative Intensity details}
\begin{tabular}{ccccccc}
\hline \\ 
Lower level & Upper level & $\lambda$ & I(line)\tablenotemark{a}  & &Center & I(fea.)\tablenotemark{b} \\
($\nu$$_{1}$$\nu$$_{2}$$\nu$$_{2}$)[JK$_{a}$K$_{c}$]  & ($\nu$$_{1}$$\nu$$_{2}$$\nu$$_{2}$)[JK$_{a}$K$_{c}$] & $\mu$m & Rel. & & $\mu$m & Rel. \\
\hline 
\hline \\ 
(0 0 0)[7 3 4]   & (0 1 1)[8 5 3]   & 1.73202 & 1.93 &\vline & 1.7329  & 1.77  \\
(0 0 0)[6 3 4]   & (0 1 1)[7 5 3]   & 1.73239 & 1.61 &\vline&        &       \\
\\
(0 0 0)[9 2 7]   & (0 1 1)[10 4 6]  & 1.73686 & 2.83 & &1.73686 & 1.93  \\
\\
(0 0 0)[15 7 8]  & (0 1 1)[16 7 9]  & 1.73987 & 1.23 &\vline	&       &       \\	
(0 0 0)[15 5 10] & (0 1 1)[16 5 11] & 1.74036 &	2.07 &\vline  &       &       \\			
(0 0 0)[16 6 11] & (0 1 1)[17 6 12] & 1.74082 &	1.03 &\vline &1.74071 & 2.21  \\
(0 0 0)[15 6 9]	 & (0 1 1)[16 6 10] & 1.74121 &	1.91 &\vline  &       &       \\			
(0 0 0)[5 3 2]   & (0 1 1)[6 5 1]   & 1.74128 &	1.26 &\vline  &       &       \\
\\			
(0 0 0)[14 8 7]	 & (0 1 1)[15 8 8]  & 1.74269 & 1.27 & &Shoulder& 1.36  \\
(0 0 0)[7 0 7]	 & (0 1 1)[8 2 6]   & 1.74445 &	1.24 &\vline	&       &       \\		
(0 0 0)[13 5 8]	 & (1 1 0)[14 6 9]  & 1.74454 &	1.23 &\vline &1.74456 & 1.77  \\
(0 0 0)[13 9 4]	 & (0 1 1)[14 9 5]  & 1.74471 &	1.12 &\vline &	       &       \\
\\		
(0 0 0)[14 7 8]	 & (0 1 1)[15 7 9]  & 1.74650 &	2.12 &\vline &	       &       \\		
(0 0 0)[6 2 5]	 & (0 1 1)[7 4 4]   & 1.74674 &	2.29 &\vline	&       &       \\		
(0 0 0)[16 5 12] & (0 1 1)[17 5 13] & 1.74680 &	1.50 &\vline &1.74688 & 2.59  \\
(0 0 0)[8 2 6]	 & (0 1 1)[9 4 5]   & 1.74728 &	1.36 &\vline	&       &       \\	 	
(0 0 0)[14 6 8]	 & (0 1 1)[15 6 9]  & 1.74750 & 1.13 &\vline	&       &       \\	 
									
\hline
\tablenotetext{a}{The relative intensities of the individual strong lines are computed at 800K}
\tablenotetext{b}{The relative intensities of  the  features are computed for 25 Jan 2003}
\end{tabular}
\end{table}

\clearpage
\begin{table}
\caption{Temperature and Column Densities}
\begin{tabular}{cccc}
\hline \\ 
Obs. & Temperature & Error & Column Density \\
Date & (K)         &  (K)  & molecules cm$^{-2}$  \\

\hline 
\hline \\ 
20 Nov 2002 & 900 & 30 & 9.3x10$^{21}$  \\
25 Jan 2003 & 800 & 30 & 9.0x10$^{21}$    \\
14 Dec 2003 & 800 & 30 & 3.8x10$^{21}$   \\
15 Apr 2004 & 800 & 40 & 5.1x10$^{21}$   \\       
25 Dec 2004 & 750 & 50 & 4.6x10$^{21}$  \\
\hline
\end{tabular} 
\end{table}

\end{document}